\documentclass[conference]{IEEEtran}
\IEEEoverridecommandlockouts
\usepackage{bm}
\usepackage{tabularx}     
\usepackage{amsmath,amssymb,amsfonts}
\usepackage{amsthm}
\usepackage{algcompatible}
\usepackage[shortlabels]{enumitem}
\usepackage[ruled]{algorithm2e}
\usepackage{algpseudocode}
\usepackage{steinmetz}
\usepackage{graphicx}
\graphicspath{{sims/}}
\usepackage{textcomp}
\usepackage{color,xcolor}
\usepackage[numbers,sort&compress]{natbib}

\newcommand{\Complex}[0]{\mathbb C}

\newcommand{\ve}[1]{{\mathbf{#1}}}

\newcommand{\mh}{\mathrm{h}}
\newcommand{\mv}{\mathrm{v}}
\newcommand{\ms}{\mathrm{s}}

\newcommand{\Herm}[0]{^{\mathrm{H}}}

\newcommand{\Trans}[0]{^{\mathrm{T}}}

\newcommand{\diag}[1]{\mbox{diag}\!\left\{#1\right\}}

\newcommand{\ee}{\mathrm{e}}

\newcommand{\jj}{\mathrm{j}}

\newtheorem{example}{Example}

\AtBeginEnvironment{example}{%
  \pushQED{\qed}%
}
\AtEndEnvironment{example}{\popQED\endexample}
\DeclareSymbolFont{symbolsC}{U}{ntxsyc}{m}{n}
\SetSymbolFont{symbolsC}{bold}{U}{ntxsyc}{b}{n}
\DeclareMathSymbol{\multimapdotbothB}{\mathrel}{symbolsC}{24}

\newif\ifShowCorrections
\ShowCorrectionsfalse
\ifShowCorrections
\usepackage[normalem]{ulem}
\definecolor{forgreen}{rgb}{0,0.6,0}
\definecolor{orange}{RGB}{255,140,0}
\definecolor{skyblue}{RGB}{100, 150, 235}

\newcommand{\fakc}[1]{{\color{purple}[#1]}}
\newcommand{\ikp}[2]{{\color{red}\sout{#1}}{\color{cyan}#2}}
\newcommand{\ikpc}[1]{{\color{forgreen}[#1]}}
\newcommand{\sw}[2]{{\color{red}\sout{#1}}{\color{blue}#2}}
\newcommand{\swc}[1]{{\color{orange}[#1]}}
\else
\newcommand{\ikp}[2]{#2}
\newcommand{\ikpc}[1]{}

\newcommand{\fakc}[1]{}
\newcommand{\sw}[2]{#2}
\newcommand{\swc}[1]{}
\fi
\algnewcommand\algorithmicfore{\textbf{for}}
\algdef{S}[FOR]{For}[1]{\algorithmicfore\ #1\ \algorithmicdo}

\usepackage{fancyhdr}

\begin{document}
\title{Decoupled Azimuth–Elevation AoA Estimation Exploiting Kronecker-Separable Steering Matrices}

\author{\IEEEauthorblockN{Faizan A.~Khattak\thanks{F.A.~Khattak is with the School of
    Computer Science, University of Leeds, Leeds LS2 9JT, UK, e-mail f.a.khattak@leeds.ac.uk.}, Ian K.~Proudler, Stephan Weiss\thanks{I.K.~Proudler and S.~Weiss are with the Dept.~of EEE, University of Strathclyde, Glasgow G1 1XW, Scotland, UK, e-mails: \{ian.proudler,stephan.weiss\}@strath.ac.uk.},  
Fazal-E Asim\thanks{F-E-Asim is with Dept. of Teleinformatics Eng., Federal University of Ceará
Fortaleza, Brazil, e-mail: fazalasim@gtel.ufc.br }.
}}

\maketitle

%
%
\begin{abstract}
Uniform rectangular arrays (URA), structured non-uniform rectangular arrays (NURA), and parallelogram-shaped (UPgA \& NUPgA) arrays admit steering vectors that can be expressed as the Kronecker product of azimuth and elevation steering vectors. Accordingly, the full steering matrix can be represented as the Khatri–Rao product of the corresponding azimuth and elevation steering matrices.
This paper exploits this structure to develop an economical subspace-decoupling framework for two-dimensional angle-of-arrival (AoA) estimation. The proposed method first extracts the joint signal subspace from the spatial covariance matrix. Then it applies a low-complexity decoupling scheme to recover the column spaces of the azimuth and elevation steering matrices. With the {estimated decoupled subspaces}, conventional one-dimensional algorithms such as MUSIC, root-MUSIC, and ESPRIT can be applied independently along each dimension, followed by pairing through a two-dimensional spectral function. Monte Carlo simulations show that the proposed approach achieves higher accuracy than state-of-the-art methods, i.e., two-dimensional MUSIC, reduced-dimension MUSIC, and two-dimensional ESPRIT, for medium- and large-scale arrays while requiring fewer snapshots, {consequently with improved spectral efficiency}. 
\end{abstract}

%
%
\section{Introduction}
Both uniform and non-uniform linear arrays have been extensively studied for angle-of-arrival (AoA) estimation, yielding a rich set of algorithms such as MUSIC, root-MUSIC, and ESPRIT and their many variants~\cite{music_orig,rootmusic,ESPRIT_orig,NULA_gridless,NULA_pdp,NULA_rootmusic}. By contrast, planar array geometries—including uniform rectangular arrays (URA)  and structured non-uniform rectangular (NURA) arrays—offer clear advantages for modern applications (MIMO~\cite{fazal24ITC}, massive MIMO~\cite{MMIMO_DOA}, 3-D localization~\cite{doa_app_1,RD_MUSIC}, etc.) but have been comparatively less explored.  Although there are AoA algorithms for URAs, e.g., reduced-dimension MUSIC (RD-MUSIC)~\cite{RD_MUSIC}, its root extraction variant root-RD-MUSIC~\cite{RD_ROOT_MUSIC}, and ESPRIT variants for MIMO radar~\cite{ESPRIT_dang,ESPRIT_dang_orig}—computationally efficient methods for structured NURAs remain scarce beyond conventional or optimized 2-D MUSIC. 
In addition, structured NURAs, which offer greater flexibility in practical deployments than URAs~\cite{SNUPA}, further motivate the development of an efficient AoA algorithm.

In this paper, we consider URA configurations and, more generally, any {array configuration} in which the full steering matrix admits a Khatri-Rao factorization (column-wise Kronecker) into two lower-dimensional steering matrices for azimuth and elevation, and introduce an efficient decoupling of the two spatial dimensions, respectively. By deriving separate signal- and noise-subspace projection matrices for each {spatial} direction, the proposed approach enables independent application of root-MUSIC, MUSIC, and ESPRIT in each spatial dimension for uniform arrays, and MUSIC or irregular root-MUSIC for structured NURAs~\cite{NULA_gridless}, at a substantially reduced order compared to existing 2-D methods. The decoupling thus yields significant computational savings while improving estimation accuracy, particularly for large arrays and in regimes limited by the number of snapshots or SNR.

%
%
\section{Signal Model}
\begin{figure}
\centering{\includegraphics[scale=0.9]{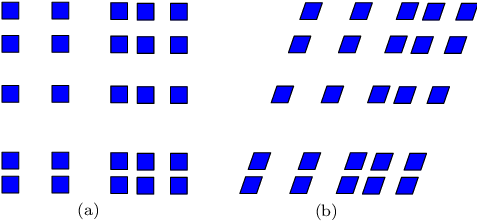}}

\vspace*{-.2cm}

\caption{A $5\times 5$ structured (a) NURA, and an (b) NUPgA array. 
The horizontal distances between sensors are row-independent, and the vertical distances between sensors are column-independent, with sensors shown as blue squares. \sw{}{Note that maximum horizontal and vertical distances between adjacent sensors do not exceed $\lambda/2$.} 
\label{fig_1}}
\end{figure}

Let us consider an $M \times N$ structured non-uniform array, where the horizontal spacing is non-uniform but row-independent, and the vertical spacing is also non-uniform but remains column-independent, as illustrated in Fig.~\ref{fig_1} for a $5 \times 5$ structured non-uniform array. 
Note that this definition encompasses the cases of a URA (both uniform and structured non-uniform) as well as a parallelogram-shaped array (both uniform, i.e., UPgA, and structured non-uniform, i.e., NUPgA). 
For simplicity, we focus on a structured NURA which includes the URA and parallelogram array.
The array is assumed to lie in the $xz$-plane, with the azimuth angle 
$\theta$ defined as the angle between the incident direction and the $x$-axis, 
and the elevation angle $\vartheta$ defined as the angle between the incident 
direction and the $z$-axis. 
Therefore, both angles are restricted to the interval 
$\theta, \vartheta \in [0,\pi]$. 
{T}he above {angle} definition follows the proposal in e.g~\cite{angle_pairing} and enables independent AoA estimation.
The steering vector {of a structured NURA} is given as a Kronecker product between horizontal and vertical steering vector 
\begin{align}
\ve{a}(\mu_{\mh},\mu_{\mv})=\ve{a}_{\mathrm{h}}(\mu_{\mh})\otimes\ve{a}_{\mathrm{v}}(\mu_{\mv})\in\mathbb{C}^{MN},
\end{align} 
where 
$\ve{a}_{\mh}(\mu_{\mh})=[1,\ee^{-\jj\mu_{\mh}p_{1,\mathrm{x}}},\dots,\ee^{-\jj  \mu_{\mh}p_{M-1,\mathrm{x}}}]$,~$\ve{a}_{\mv}(\mu_{\mv})=[1,\ee^{-\jj\mu_{\mv}p_{1,\mathrm{z}}},\dots,\ee^{-\jj \mu_{\mv}p_{N-1,\mathrm{z}}}]\Trans$, $p_{m,\mathrm{x}}$ and $p_{n,\mathrm{z}}$ are sensor positions along $x$ and $z$ axis with $p_{0,\mathrm{x}}=0,p_{0,\mathrm{z}}=0$. 
The operator $\otimes$ denotes the Kronecker product, while $\mu_{\mh}$
and $\mu_{\mv}$
are horizontal and vertical spatial frequencies, respectively, of {a source}.
For $P$ narrowband sources $\ve{x}(t)\in\mathbb{C}^{P}$ in far-field, the sensor measurement at discrete time index $n$ \sw{}{is} 
\begin{align}
\ve{y}[n]=\ve{A}\ve{x}[n]+\ve{w}[n]\sw{,}{} {\in\mathbb{C}^{MN}} \sw{}{\; ,}
\end{align}
where
\begin{align}
\ve{A}&=[\ve{a}_{\mathrm{h}}(\mu_{\mh,1})\otimes\ve{a}_{\mathrm{v}}(\mu_{\mv,1}),\dots,\ve{a}_{\mathrm{h}}(\mu_{\mh,P})\otimes\ve{a}_{\mathrm{v}}(\mu_{\mv,P})]\nonumber\\
&=\ve{A}_{\mh}\diamond\ve{A}_{\mv}\in\mathbb{C}^{MN\times P}
\end{align} 
is the steering matrix and can be expressed as a Khatri-Rao product, denoted by $\diamond$, between steering matrices in azimuth and elevation direction, and $\ve{w}[n]\in\mathbb{C}^{MN}$ represents a noise term.
We further assume that the source spatial covariance matrix is of full rank $P$ which includes the case of uncorrelated sources and coherent sources with spatial smoothing, and noise is white with zero mean and variance $\sigma_{{w}}^2$. With noise mutually uncorrelated with {the} sources, we have 
\begin{align}
\ve{R}=\ve{A}\ve{R}_{\mathrm{xx}}\ve{A}\Herm+\sigma_{{w}}^2\ve{I}_{MN\times MN}\; ,
\end{align}
where $\ve{R}_{\mathrm{xx}}$ is the source covariance matrix.
\section{Subspace Recovery of Khatri-Rao Factors from Their Product Under Invertible Transformation}
{Both} MUSIC and ESPRIT algorithms require knowledge of the {noise} and signal subspace, respectively. 
In the following, we show how to estimate the azimuth-only and elevation-only signal subspaces.

\subsection{Subspace Information In Kronecker Structure}
Via an {eigenvalue decomposition} of $\ve{R}$, we have
\begin{align}
\ve{R}=\ve{Q}_s\ve{\Lambda}_{\ms}\ve{Q}_{\ms}\Herm+\ve{Q}_{\perp}\ve{\Lambda}_{\perp}\ve{Q}_{\perp}\Herm,
\end{align}
where $\ve{Q}_{\ms}\in\mathbb{C}^{MN\times P}$ contains the {eigenvectors{ that span} the signal-only subspace,  corresponding to the} $P${} dominant eigenvalues in $\ve{\Lambda}_{\ms}$, whereas the {remaining} eigenvector{s} in $\ve{Q}_{\perp}$ span the noise-only subspace.
With $\ve{R}_{\mathrm{xx}}$ full rank, $\mathrm{span}(\ve{A})=\mathrm{span}(\ve{Q}_{\ms})$ {is guaranteed}~\cite{music_orig}, and so we have 
\begin{align}
\ve{Q}_{\ms}=\ve{A}\ve{G}
  \sw{}{\; ,}
\end{align} 
where $\ve{G}\in\mathbb{C}^{P\times P}$ is an invertible matrix~\cite{basis}. 
This serves as the foundation for {our} subspace {} {decomposition approach}.
With $\ve{A}$ admitting {a} Khatri-Rao factorization, the \( p \)th column of $\ve{Q}_{\ms}$ is given as
\begin{align}
\ve{q}_p=\sum_{k=1}^{P}g_{k,p}\ve{a}_{\mh,k}\otimes\ve{a}_{\mv,k},
\end{align}
where $g_{k,p}$ is the element at $k$th row and $p$th column of $\ve{G}$.
With $\ve{q}_{p}{\in \Complex^{NM}}$ reshaped into an \( N \times M \) matrix \( \ve{Q}_{\mathrm{s},p} \) via an unvec operation, we have  
\begin{align}
\label{eq_term_1}
\ve{Q}_{\ms,p}=\sum_{k=1}^{P}g_{k,p}\ve{a}_{\mv,k}\ve{a}_{\mh,k}\Trans=\ve{A}_{\mv}\diag{\ve{g}_p}\ve{A}_{\mh}\Trans,
\end{align}
It is tempting to think that the column space of $\ve{Q}_{\ms,p}$ spans the azimuth-only subspace, and that the row space of $\ve{Q}_{\ms,p}$ spans the elevation-only subspace. 
However, this is not the case as $\ve{G}$ may contain some zero elements.
Therefore, a reformulation is necessary, outlined in the next subsection, in order to extract the column spaces of both $\ve{A}_\mh$ and $\ve{A}_\mv$. 

\subsection{Azimuth and Elevation Column Spaces Retrieval }

In order to access the column spaces of $\ve{A}_{\mv}$ and $\ve{A}_{\mh}$, we stack 
\begin{align}
\label{eq_C}
\ve{C}_r=&[\{\ve{Q}_{\ms,1}\}_r,\dots,\{\ve{Q}_{\ms,P}\}_r] \in \mathbb{C}^{N\times P}\nonumber\\
=&[\ve{A}_{\mv}\diag{\ve{g}_1}\{\ve{A}_{\mh}\Trans\}_k,\dots,\ve{A}_{\mv}\diag{\ve{g}_P}\{\ve{A}_{\mh}\Trans\}_r]\nonumber\\
=&\ve{A}_{\mv}\ve{D}_{\mh,r}\ve{G},
\end{align} 
where $\ve{D}_{\mh,r}=\diag{\{\ve{A}_{\mh}\Trans\}_r}$\ikp{}{, and $\{\ve{F}\}_r$ is the $r$th column of a matrix $\ve{F}$}. 
Since none of the element of $\ve{A}_{\mh}$ are zero, $\ve{C}_{r}$ is full column rank for every value of $r$. Therefore, one can compute the column space of $\ve{A}_{\mv}$ from $\ve{C}_r$, however it will not be very accurate because not all of the available data is being used. Instead we concatenate $\ve{C}_r$ for $r=1,\dots,P$ horizontally as
\begin{align}
\label{eq_term_C1}
\ve{C}&=[\ve{C}_1,\dots,\ve{C}_{M}]=[\ve{A}_{\mv}\ve{D}_{\mh,1}\ve{G} \dots \ve{A}_{\mv}\ve{D}_{\mh,M}\ve{G}]\nonumber\\
&=\ve{A}_{\mv}[\ve{D}_{\mh,1}\ve{G} \dots \ve{D}_{\mh,M}\ve{G}] \in \mathbb{C}^{N\times MP}.
\end{align}
With each block $\ve{D}_{\mh,m}\ve{G}\in\mathbb{C}^{P\times P},~m=1,\dots,M$ invertible, the column space of $\ve{C}$ is {the} same {a}s that of $\ve{A}_{\mv}$, and therefore the orthonormal basis for the column space can be obtained from the SVD of $\ve{C}$ and collecting $P$ dominant left singular vectors of $\ve{C}$ as
\begin{align}
\mathrm{col}\{\ve{A}_{\mv}\}=\mathrm{span}\{\ve{u}_{\mathrm{C},1},\dots,\ve{u}_{\mathrm{C},P}\} \sw{}{\; ,}
   \label{eqn:colAh}
\end{align} 
where $\ve{u}_{\mathrm{C,n}}$ is the $n$th dominant left singular vector of $\ve{C}$. Similarly, stacking the $r$th row of $\ve{Q}_{\ms,p}$ for $p=1,\dots,P$, we obtain
\begin{align*}
\ve{B}_r=[\{\ve{Q}_{\ms,1}\Trans\}_r,\dots,\{\ve{Q}_{\ms,P}\Trans\}_r]=\ve{A}_{\mh}\ve{D}_{\mv,r}\ve{G}\in\mathbb{C}^{M\times P} \; ,
\end{align*}
where~$\ve{D}_{\mv,r}=\diag{\{\ve{A}_{\mv}\Trans\}_r}$, and concatenating horizontally to construct
\begin{align}
\label{eq_term_C1a}
\ve{B}&=[\ve{B}_1,\dots,\ve{B}_{N}]=[\ve{A}_{\mh}\ve{D}_{\mv,1}\ve{G} \dots \ve{A}_{\mh}\ve{D}_{\mv,N}\ve{G}]\nonumber\\
&=\ve{A}_{\mh}[\ve{D}_{\mv,1}\ve{G} \dots \ve{D}_{\mv,N}\ve{G}]\in\mathbb{C}^{M\times NP} \; .
\end{align}
The set of orthonormal vectors spanning the column space of $\ve{A}_{\mh}$ can similarly be obtained as
\begin{align}
\mathrm{col}\{\ve{A}_{\mh}\}= \mathrm{span}\{\ve{u}_{\mathrm{B},1},\dots,\ve{u}_{\mathrm{B},P}\}   \sw{}{\; ,}
   \label{eqn:colAv}
\end{align} 
where $\ve{u}_{\mathrm{B,n}}$ is the $n$th dominant left singular vector of $\ve{B}$. 

With \eqref{eqn:colAh} and \eqref{eqn:colAv}, we now have orthonormal bases for the column spaces of both $\ve{A}_\mh$ and $\ve{A}_\mv$.
It can be seen that with this decoupling approach, only $\mathrm{min}\{M,N\}-1$ many sources can be detected, which is less than other approaches, i.e., $MN-1$ sources by 2D-MUSIC.

\subsection{Decoupled Azimuth and Elevation Spatial Frequencies Estimation}
\subsubsection{Uniform Rectangular or Parallelogram Arrays }
With an orthonormal basis set for the column space of $\ve{A}_\mh$ and $\ve{A}_{\mv}$ obtained from the above decoupling approach, {a} projection matrix onto the noise subspace can be obtained. So all subspace based AoA estimation methods applicable to a ULA such as MUSIC~\cite{music_orig}, root-MUSIC~\cite{rootmusic}, gold-MUSIC~\cite{gold_music}, and ESPRIT~\cite{ESPRIT_orig} can be applied to determine the horizontal and vertical spatial frequencies of any source contributions independently. 
In this paper, we refer to applying root-MUSIC and ESPRIT after obtaining the column space in each direction as decoupled root-MUSIC (De-RMUSIC) and decoupled ESPRIT (De-ESPRIT), respectively.
\sw{}{It is important to note that the proposed method is a purely matrix-based technique rather than a tensor-based approach such as tensor-MUSIC or tensor-ESPRIT \cite{tensorMUSIC}, and therefore is inherently less computationally expensive. This also contrasts with recently proposed methods based on parallel factorization and a tensor SVD \cite{ParafacDoA}.}

\subsubsection{Structured Non-uniform Rectangular or Parallelogram Arrays}
With the proposed decoupling approach, once the orthonormal set of basis vectors are obtained, the conventional MUSIC or the irregular Vandermonde decomposition-based root-MUSIC algorithm~\cite{NULA_gridless} can be applied in each direction to estimate spatial frequencies both for structured NURA and NUPgAs.
Due to this decoupling, $n_\mh \times n_\mv$ grid searches required in the 2D-MUSIC reduces to only $n_\mh + n_\mv$. 
For example, with $1^\circ$ resolution over a scanning range of $-90^\circ$ to $90^\circ$, we have $n_h = n_v = 181$, giving $32{,}761$ searches for 2D-MUSIC, but only $362$ searches with the decoupled method, achieving a $90.5$-fold reduction in computational cost.

\subsection{Angle Pairing}
To correctly pair an estimated azimuth angle with its corresponding elevation angle, here we rely on the conventional method~\cite{RD_ROOT_MUSIC} of evaluating {at all pairs of $a_h$ and $a_v$} the pseudospectrum 
\begin{align}
P_{\mathrm{MUSIC}}(\mu_\mh,\mu_\mv)=\ve{a}(\mu_{\mh},\mu_{\mv})\Herm\ve{Q}_{\perp}\ve{Q}_{\perp}\Herm\ve{a}(\mu_{\mh},\mu_{\mv})   \sw{}{\; ,} 
\end{align}
and perform the associations through finding peaks. 
{This pairing makes sure that estimate azimuth and elevation angles correspond to the same source}

\section{Results and Simulation}

\subsection{URA Case}
We simulate a scenario with {$M\times M$} URA\sw{}{s} of size {$M \in\{10,20,30\}$}, and \( P = 3 \) sources located at 
\( (\theta_1, \vartheta_1) = (155^\circ, 20^\circ) \), \( (\theta_2, \vartheta_2) = (21^\circ, 150^\circ) \), and \( (\theta_3, \vartheta_3) = (76^\circ, 80^\circ) \). 
The covariance matrix $\ve{R}$ is estimated from \(L=100\) snapshots across various signal-to-noise ratios (SNRs). 
Performance is evaluated using the root mean square error (RMSE), defined as  
\begin{align}
\zeta=\frac{1}{I}\sum_{i=1}^{I}\sqrt{\frac{1}{P}\sum_{p=1}^{P}(\theta_{p}-\hat{\theta}_{p,i})^2+(\vartheta_{p}-\hat{\vartheta}_{p,i})^2} \sw{}{\; ,}
\end{align}  
where \( \hat{\theta}_{p,i}, \hat{\vartheta}_{p,i} \) denote the \( p \)th estimated angles in the \( i \)th Monte Carlo run for azimuth and elevation direction, respectively, and $I$ is the number of Monte-Carlo runs.
For each {SNR} value{} we simulate $I=5000$ runs.

\begin{figure}
\centering{\includegraphics[width=\columnwidth]{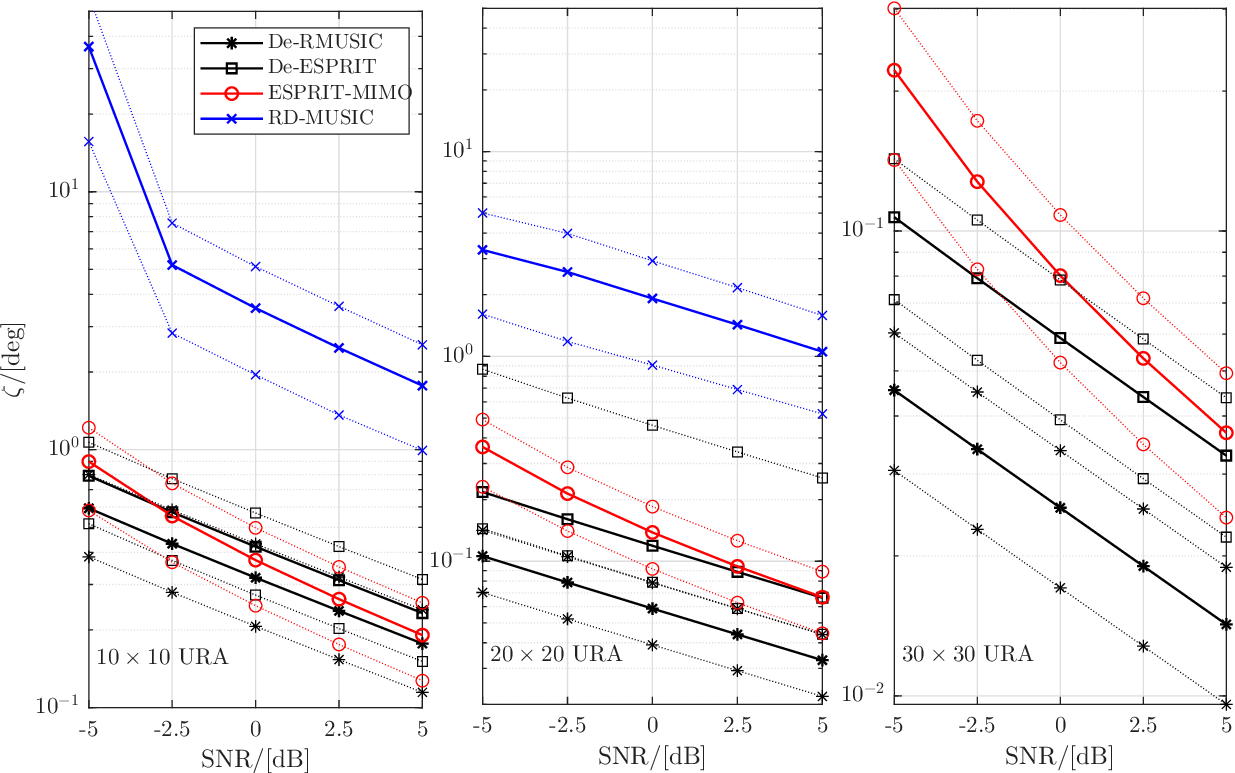}}

\caption{Ensemble simulation showing $\zeta$ versus $\mathrm{SNR}$ for $L=100$ snapshots for URA of various sizes\sw{.}{} (bold line is mean and dotted line is mean $\pm$ standard deviation)\sw{}{.}}
\label{fig_RMSE}
\end{figure}

We compare the proposed De-RMUSIC and De-ESPRIT against RD-MUSIC~\cite{RD_MUSIC}, simulated at a resolution of \( 0.05^\circ \), and the pairing-free ESPRIT method~\cite{ESPRIT_dang}, proposed originally for MIMO radar, which is directly applicable to URAs.
\sw{}{In this comparison, we do not consider tensor-MUSIC~\cite{tensorMUSIC} as the proposed method is an ordinary matrix manipulation based angle of arrival estimation. For the same reason, we exclude the tensor SVD~\cite{ParafacDoA} from our comparison.} 
The ensemble results are shown in Fig.~\ref{fig_RMSE}, 
where, for the $10 \times 10$ size URA, De-RMUSIC, ESPRIT-MIMO and De-ESPRIT
are statistically indistinguishable except at $-5\mathrm{dB}$ SNR where De-RMUSIC 
performs the best. RD-MUSIC clearly has the worst performance. 
For the larger size URAs, i.e. $20 \times 20$ and $30 \times 30$, De-RMUSIC consistently {performs best} at all SNR values. 
\sw{}{One apparent reason for the higher accuracy of De-RMUSIC is that it exploits the complete data set when estimating both the azimuth and elevation parameters, rather than partitioning the data into two overlapping subsets for parameter estimation, as is done in ESPRIT-based methods.}
This is particularly clear for 
the $30 \times 30$ array. ESPRIT-MIMO and De-ESPRIT
are again mostly statistically indistinguishable. An exception is for the 
30 × 30 array at low SNRs, where De-ESPRIT outperforms ESPRIT-MIMO. 
For the $20 \times 20$ array RD-MUSIC has again the worst performance. 
As such, we exclude it from the remaining comparison.

\begin{figure}
\centering{\includegraphics[width=\columnwidth]{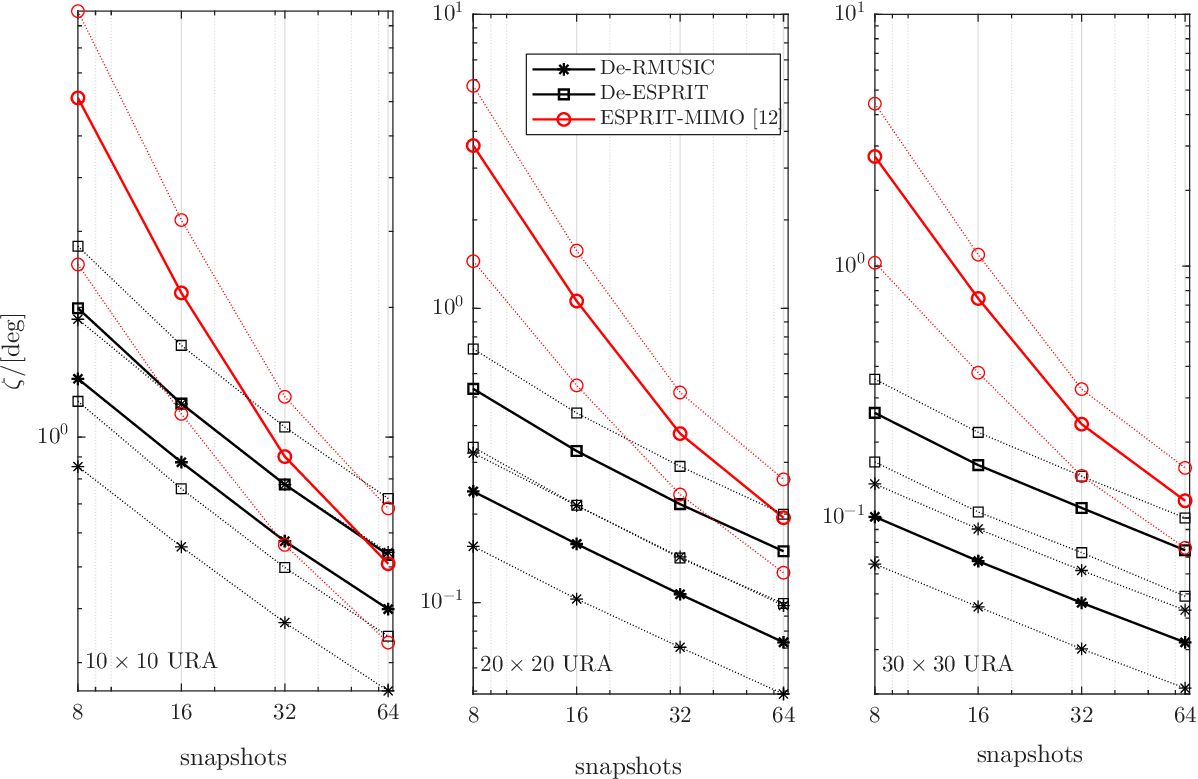}}

\caption{{RMSE} $\zeta$ vs number of snapshots for URA at $\mathrm{SNR}=0\mathrm{dB}$ for $P=3$ sources. The solid line is the mean, and the dotted line is the mean $\pm$ standard deviation.
\label{fig_snapshots}}
\end{figure}

To evaluate the performance of the proposed De-RMUSIC and De-ESPRIT algorithms with varying numbers of snapshots used for spatial covariance matrix estimation, we measure the RMSE at $\mathrm{SNR}=0\ \mathrm{dB}$ for $L=2^\ell,\ \ell=3,4,5,6$, for the considered URA sizes. For each value of $L$, 5000 Monte Carlo runs are performed. The resulting performance trends are illustrated in~Fig.~\ref{fig_snapshots}, showing that both proposed algorithms
consistently outperforms the existing ESPRIT-MIMO  below a certain number of snapshots.  Above this, the algorithms are statistically indistinguishable. The threshold {depends} on the size of the array{}.
These results highlight the superior accuracy of the proposed approaches, especially under low-SNR and limited-snapshot conditions, where ESPRIT-MIMO exhibits significantly degraded performance.

\begin{figure}
    \centering{\includegraphics[width=\columnwidth]{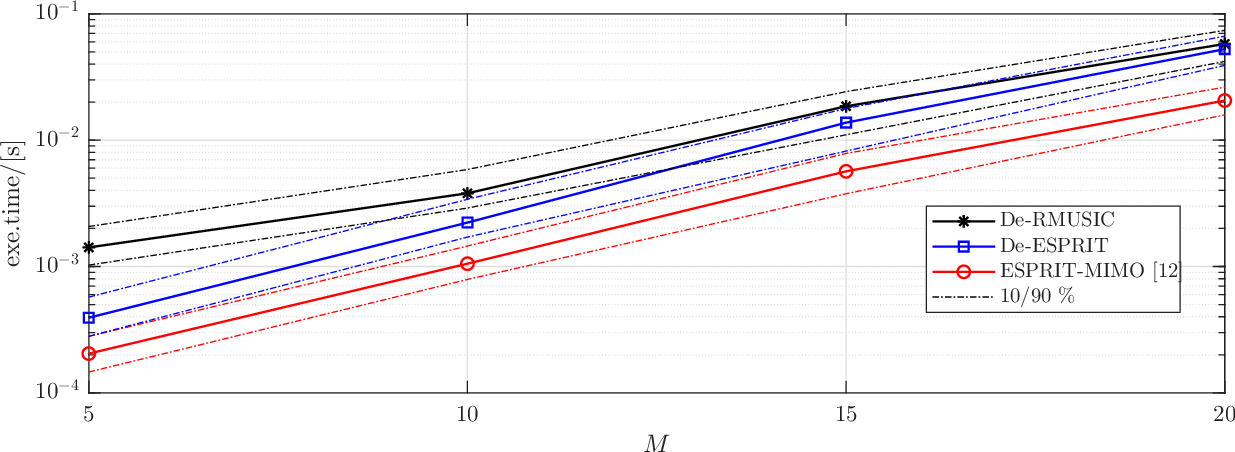}}

\caption{Execution time vs $M$ where $M=N$ for simulating $P=(M-1)$ sources for the URA{;} solid lines denote mean{,} dotted line{s percentiles.}
}
\label{fig_time}
\end{figure}

It is evident that the proposed algorithms are computationally more efficient than RD-MUSIC, yet slightly more expensive than the ESPRIT-MIMO algorithm due to the additional angle pairing and the two extra SVDs of $\mathbf{B}$ and $\mathbf{C}$. Nevertheless, given that the proposed algorithms achieve superior accuracy in both low-SNR and small-snapshot scenarios for large arrays, we evaluate the execution time in the MATLAB environment for $P = (M-1)$ on an $M \times M$ size arrays with $M = 5, 10, 15, 20$.
The resulting execution time for competing algorithms are shown in Fig.~\ref{fig_time}. 
The RD-MUSIC relies on a line search and is therefore computationally expensive; hence, it is not included in this comparison.
The ESPRIT-MIMO is consistently superior to the two proposed approaches, due to being pairing-free, with the De-ESPRIT being the second best. 
As evident, the performance gain in computation time of ESPRIT-MIMO over the proposed approaches is not significant over shown spatial dimension but may become apparent at even higher spatial dimensions. 
Therefore, applications requiring higher accuracy with fewer snapshots at low SNR conditions may prefer the proposed methods. 
Furthermore, the computational time gap with ESPRIT-MIMO can be further reduced by combining the gold-MUSIC, reported to exhibit lower complexity than root-MUSIC~\cite{gold_music}, with the proposed decoupling approach.


\subsection{Structured NURA Case}
{W}e {next} compare the proposed decoupling approach combined with independent spectral search MUSIC in each direction, referred to as De-MUSIC, against the conventional 2D spectral-search MUSIC algorithm. 
The comparison considers $P=3$ sources at same azimuth and elevation angles as {in the above} experiment but the URA is replaced with a structured NURA of sizes $10\times 10$ and $20\times 20$. 
Both MUSIC and 2D-MUSIC~ employ a two-stage search{}: an initial coarse search with $1^\circ$ resolution followed by a fine search with $0.05^\circ$ resolution around the approximate AoAs obtained from the coarse search. 
Alternatively, the fine search is replaced by a single-variable nonlinear optimization using MATLAB's \texttt{fminbnd} function, with bounds derived from the coarse search results. 
This variant is referred to as De-MUSIC-Opt.

\begin{figure}
\centering{\includegraphics[width=\columnwidth]{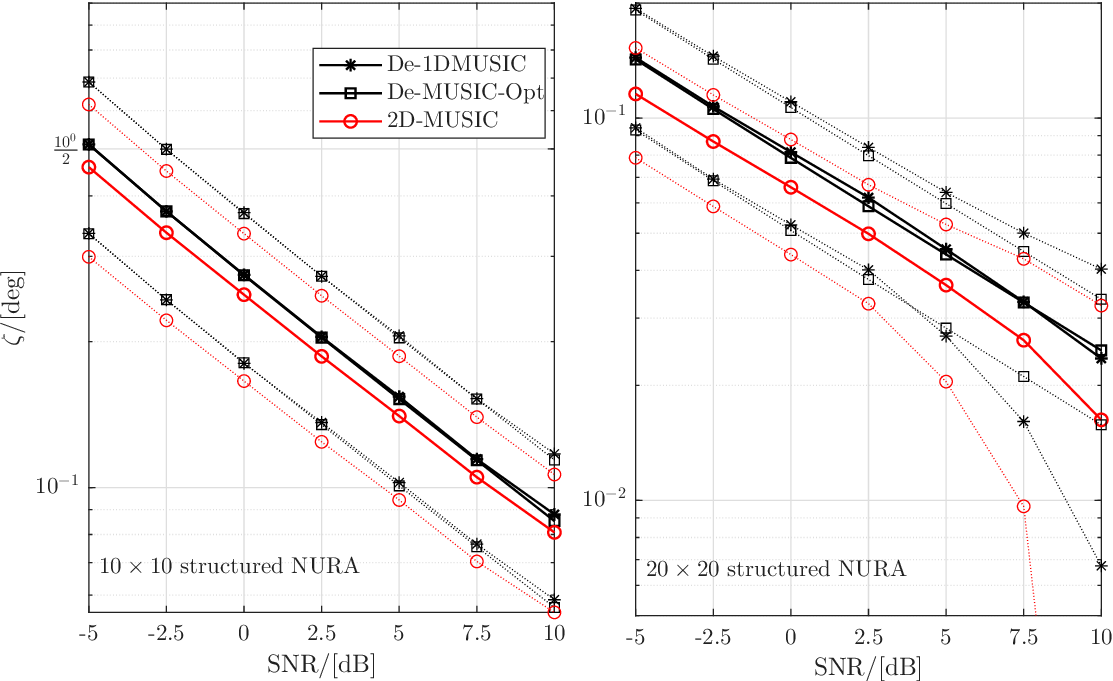}}

\vspace*{-.3cm}

\caption{{RMSE} $\zeta$ vs SNR for $L=100$ snapshots with two sizes {of} structured NURAs{;} solid line{s} denote{} mean, dotted line {} mean $\pm$ standard deviation){.}}
\label{fig_NURA_1}
\end{figure}

\begin{figure}
\centering{\includegraphics[width=\columnwidth]{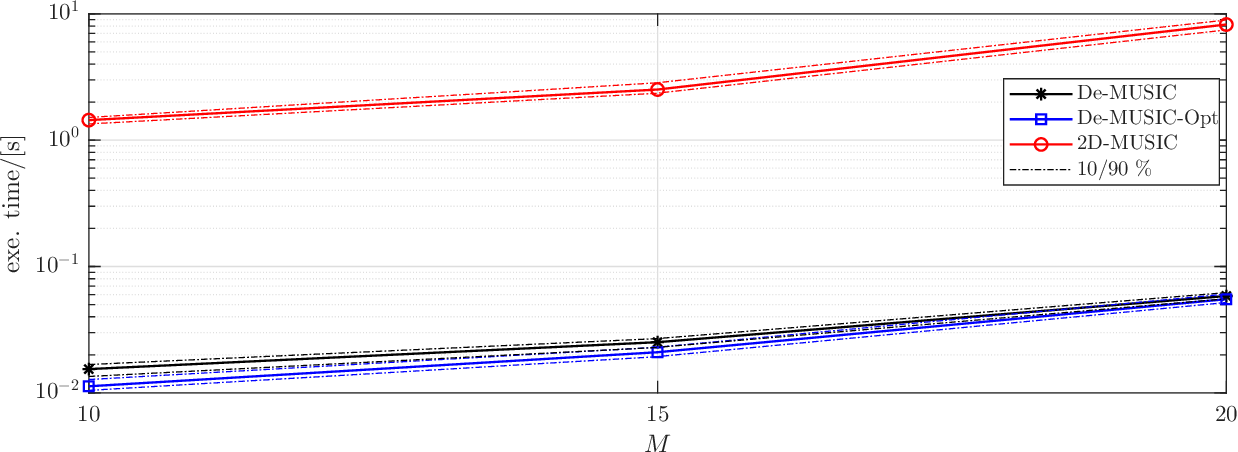}}

\caption{Execution time vs $M$ where $M=N$ for $P=3$ sources for the structured NURA{;} (solid lines denote mean{,} dotted line{s} {percentiles}.
\label{fig_NURA_2}}
\end{figure}

\begin{figure}
\centering{\includegraphics[width=\columnwidth]{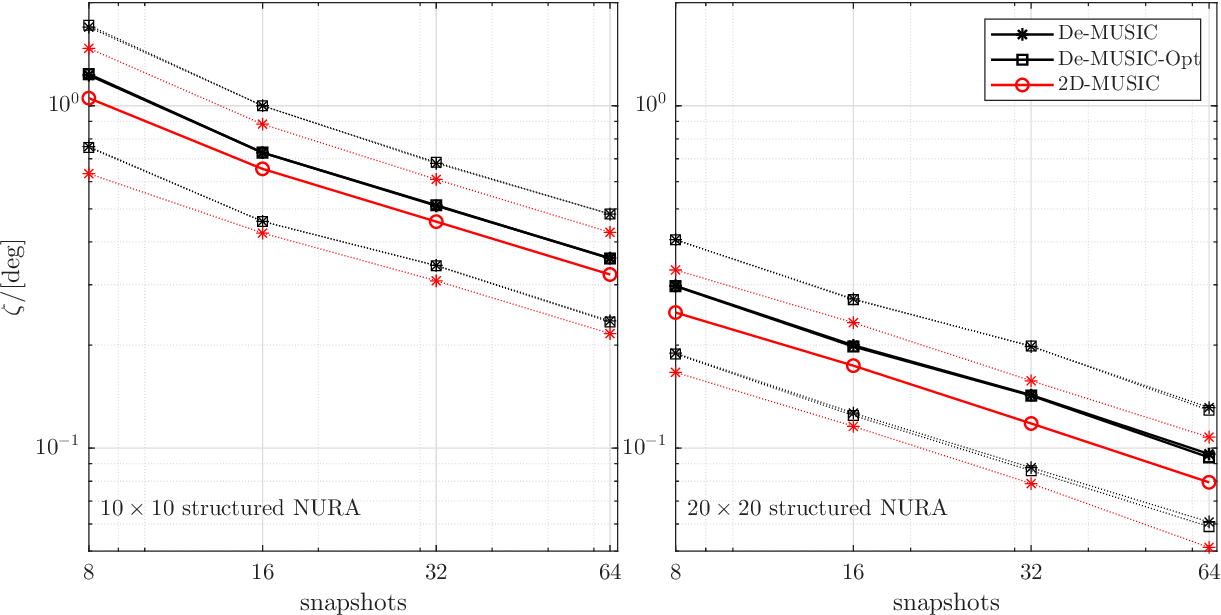}}

\caption{Ensemble simulation of $\zeta$ versus the number of snapshots for 
structured NURA at $\mathrm{SNR}=0\,\mathrm{dB}$ with $P=3$ sources. Bold lines 
denote the mean, while dotted lines indicate the mean $\pm$ one standard deviation 
(black: $10 \times 10$ NURA, red: $20 \times 20$ NURA).}
\label{fig_NURA_3}
\end{figure}

The RMSE versus SNR profiles in Fig.~\ref{fig_NURA_1} demonstrate that both proposed approaches achieve accuracy comparable to 2D-MUSIC while being significantly more computationally efficient, as illustrated in Fig.~\ref{fig_NURA_2} with De-MUSIC-Opt being the most efficient. 
Furthermore, its performance remains robust even with a smaller number of snapshots (see Fig.~\ref{fig_NURA_3}), combining the high accuracy of 2D-MUSIC with the computational efficiency of MUSIC and maintaining effectiveness under limited snapshot scenarios.

\section{Conclusion}
We have presented a decoupling strategy that separates azimuth and elevation components in array configurations where the overall steering matrix admits a Khatri–Rao product structure between two lower-dimensional steering matrices . 
This decoupling approach is applicable in URAs/UPgAs, MIMO radar configurations, and structured NURAs/NUPgAs~\cite{SNUPA}. 
\sw{}{It allows classical subspace-based algorithms, such as MUSIC, root-MUSIC, and ESPRIT, to be applied independently in each spatial dimension instead of relying on tensor based methods which can be computationally expensive compared to ordinary matrix based methods.} 
{While the suggested approach can only deal with a limited number of sources, o}ur results show that for large uniform arrays, combining root-MUSIC or ESPRIT with the proposed decoupling achieves high accuracy across a wide SNR range—even with a limited number of snapshots. Despite the need for a pairing step to associate azimuth and elevation estimates, the proposed algorithms are computationally competitive with that of~\cite{ESPRIT_dang}.
Moreover, for structured NURAs, the approach enables independent spectral search MUSIC in each dimension, significantly reducing computational complexity while maintaining accuracy comparable to 2D-MUSIC.

\bibliographystyle{IEEEtran}
\footnotesize{
\bibliography{IEEEabrv,bib_total}
}

\end{document}